\documentclass[lettersize,journal]{IEEEtran}
\usepackage{amsmath,amsfonts}
\usepackage{algorithmic}
\usepackage{algorithm}
\usepackage{array}
\usepackage[caption=false,font=footnotesize,labelfont=rm,textfont=rm]{subfig}
\usepackage{textcomp}
\usepackage{stfloats}
\usepackage{url}
\usepackage{verbatim}
\usepackage{graphicx}
\usepackage{cite}
\usepackage{multirow}
\usepackage{booktabs}
\usepackage[numbers,sort&compress]{natbib}
\usepackage{hyperref}

\begin{document}

\title{An Efficient Implicit Neural Representation Image Codec Based on Mixed Autoregressive Model for Low-Complexity Decoding}

\author{
Xiang Liu, Jiahong Chen, Bin Chen, Zimo Liu, Baoyi An, Shu-Tao Xia and  Zhi Wang
\thanks{Xiang Liu, Jiahong Chen, Shu-Tao Xia and Zhi Wang are with the Department of Data and Information, Shenzhen International Graduate School, Tsinghua University, Beijing 100084, China (e-mail: \href{mailto:liuxiang22@mails.tsinghua.edu.cn}{liuxiang22@mails.tsinghua.edu.cn}; \href{mailto:jh-chen22@mails.tsinghua.edu.cn}{jh-chen22@mails.tsinghua.edu.cn};  \href{mailto:xiast@sz.tsinghua.edu.cn}{xiast@sz.tsinghua.edu.cn}; \href{mailto:wangzhi@sz.tsinghua.edu.cn}{wangzhi@sz.tsinghua.edu.cn}), Xiang Liu and Shu-Tao Xia are with Peng Chen Laboratory, Shenzhen China.
}
\thanks{Bin Chen is with Harbin Institute of Technology, Shenzhen 518055, China (e-mail: \href{mailto:chenbin2021@hit.edu.cn}{chenbin2021@hit.edu.cn})
}
\thanks{Zimo Liu is with the Peng Chen Laboratory, Shenzhen, China (e-mail: \href{mailto:liuzm@pcl.ac.cn}{liuzm@pcl.ac.cn}) }
\thanks{Baoyi An is with the Huawei Technology, Shenzhen, China (e-mail: \href{mailto:anbaoyi@huawei.com}{anbaoyi@huawei.com})}
}


\markboth{}
{Shell \MakeLowercase{\textit{et al.}}: A Sample Article Using IEEEtran.cls for IEEE Journals}
\maketitle

\begin{abstract}
Displaying high-quality images on edge devices, such as augmented reality devices, is essential for enhancing the user experience. However, these devices often face power consumption and computing resource limitations, making it challenging to apply many deep learning-based image compression algorithms in this field. Implicit Neural Representation (INR) for image compression is an emerging technology that offers two key benefits compared to cutting-edge autoencoder models: low computational complexity and parameter-free decoding. It also outperforms many traditional and early neural compression methods in terms of quality. In this study, we introduce a new Mixed AutoRegressive Model (MARM) to significantly reduce the decoding time for the current INR codec, along with a new synthesis network to enhance reconstruction quality. MARM includes our proposed AutoRegressive Upsampler (ARU) blocks, which are highly computationally efficient, and ARM from previous work to balance decoding time and reconstruction quality. We also propose enhancing ARU's performance using a checkerboard two-stage decoding strategy. Moreover, the ratio of different modules can be adjusted to maintain a balance between quality and speed. Comprehensive experiments demonstrate that our method significantly improves computational efficiency while preserving image quality. With different parameter settings, our method can achieve over a magnitude acceleration in decoding time without industrial level optimization, or achieve state-of-the-art reconstruction quality compared with other INR codecs. To the best of our knowledge, our method is the first INR-based codec comparable with \citet{Ball2018Variational} in both decoding speed and quality while maintaining low complexity.
\end{abstract}


\section{Introduction}

\IEEEPARstart{R}{ecent} years have seen dramatic advancements in deep learning-based lossy image compression \cite{balle2016end, Ball2018Variational, 2018Joint, 2022ELIC}. These advancements have led to significant progress, outperforming many traditional image codecs such as JPEG \cite{wallace1992jpeg} and BPG \cite{bpg} across common metrics like PSNR and MS-SSIM \cite{wang2003multiscale}. Joint backward-and-forward adaptive entropy modeling is a crucial technique in these models, utilizing side information in forward adaptation and predictions from the causal context of each symbol in backward adaptation \cite{minnen2018image, 2020Channel, cheng2020learned}.
In addition to neural image codecs based on autoencoders (AE), there has been an emergence of implicit neural representation (INR) in 3D applications, utilizing neural network weights to represent information, and this has spurred exploration into similar technologies in image compression. \citet{dupont2021coin} proposed using 2D coordinates as the input for the MLP and directly outputting the RGB value of the corresponding pixel. Expanding on this, \citet{ladune2022cool} introduced COOL-CHIC, which utilizes trainable latent variables as the input for the MLP.

\begin{figure}[t]
\centering
		\centering
        \includegraphics[width=\linewidth, trim={2cm 2.6cm 2.1cm 2.5cm },clip]{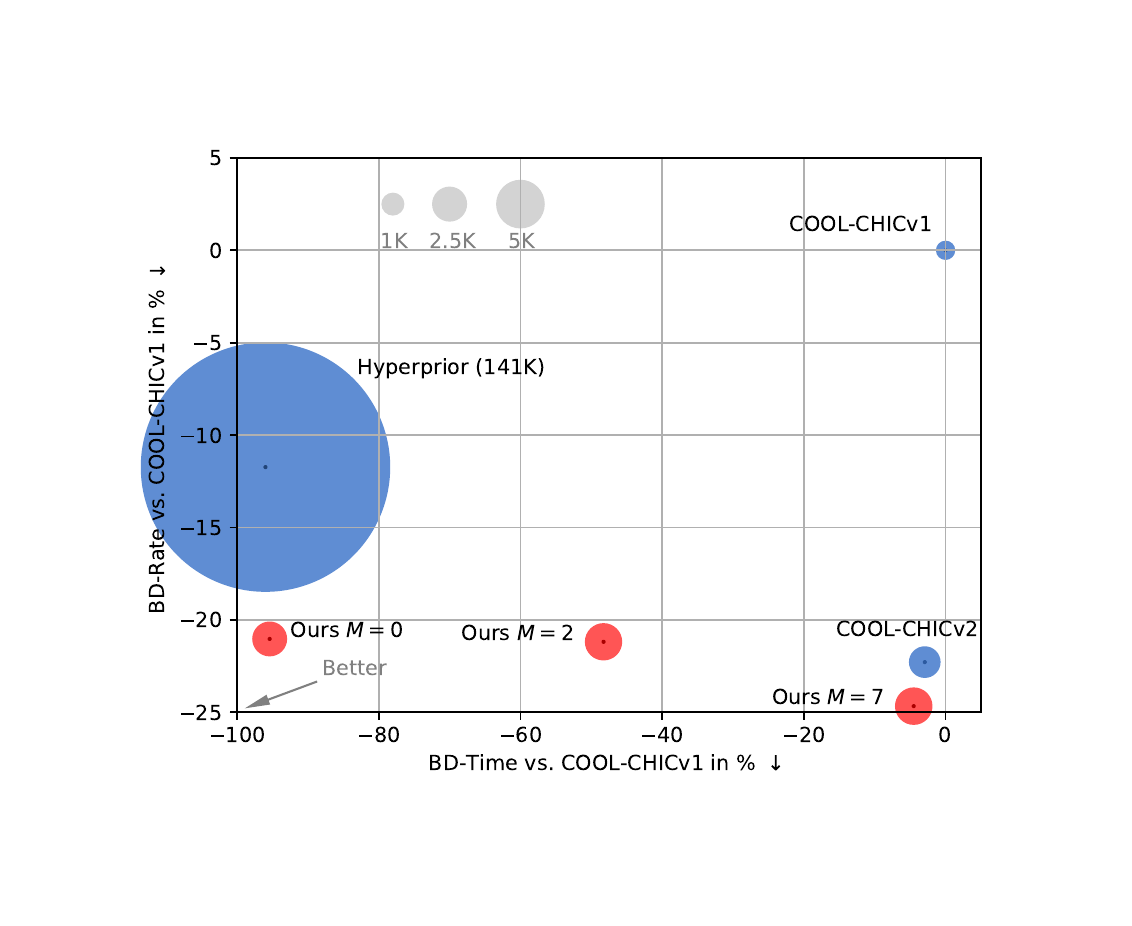}
        \caption{Overall performance of different methods on CLIC dataset. The area represents the decoding complexity in MACs/pixel (small is better). BD-Time is a metric measuring the decoding speed among different qualities like BD-Rate, which we will discuss more in Section \ref{sec:exp:subsec1}. Note on parallel accelerators like GPU, low complexity does not necessarily mean fast decoding and vice versa (\textit{e.g.} Hyperprior \cite{Ball2018Variational} and COOL-CHIC \cite{ladune2022cool}). Our method achieves a better trade-off than existing methods}
        \label{fig:header_plot}
\end{figure}

\begin{figure}[t]
\centering
		\centering
        \includegraphics[width=\linewidth, trim={0 0.0 0 0.0 },clip]{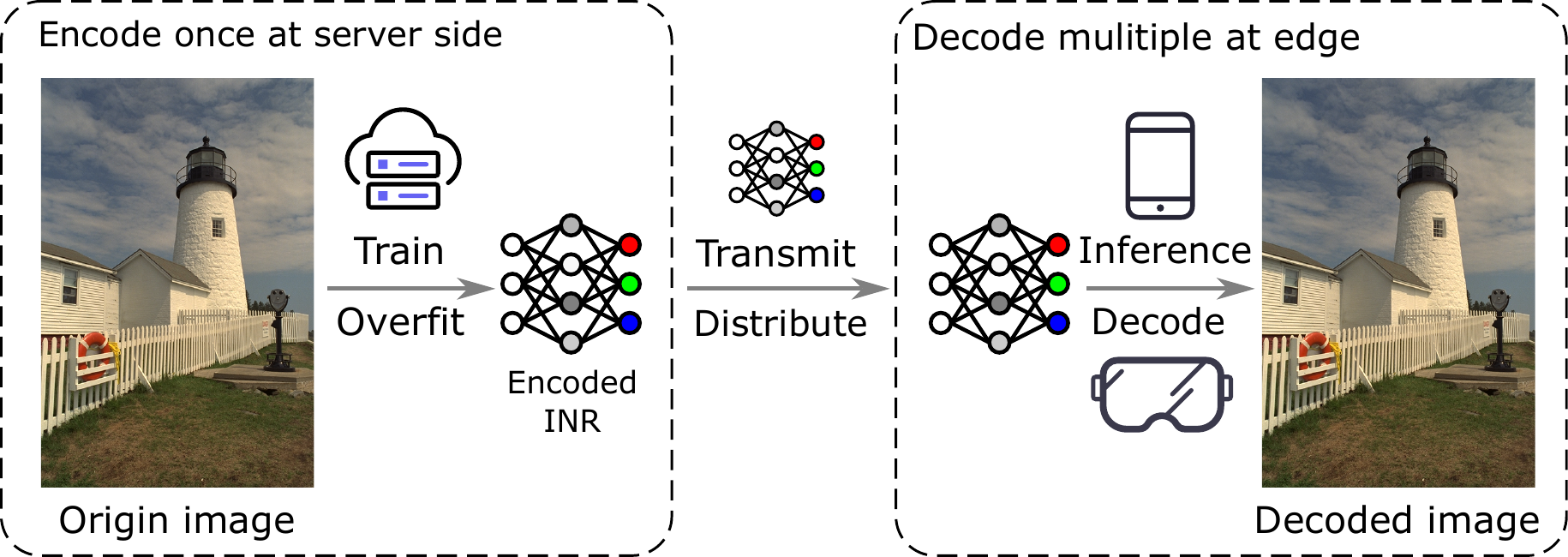}
        \caption{Asymmetric compression and decompression framework of the INR codec. }
        \label{fig:asym_arch}
\end{figure}

Although AE-based methods have achieved better rate-distortion performance, INR-based methods offer several advantages. The most impressive characteristic is INR providing a low-complexity decoding method, which is essential for AR/VR devices and low-power devices such as smartphones or AR glasses.
According to \citet{leguay2023low}, INR-based methods achieve a similar BD-Rate as HEVC and nearly two orders of magnitude less MACs (Multiplication-Accumulation) compared with AE-based methods. Moreover, unlike AE-based models, the decompression process of the INR codec does not require model parameters other than the transmitted part, resulting in a more lightweight decoder. This lightweight decoder can also easily achieve backward compatibility because there is no need to save all versions of model weights in the decoder. One challenge of INR codec is its time-consuming encoding process, which equivalent to training a small neural network. But this drawback can be alleviated.
In the context of an asymmetric compression system, as shown in Fig. \ref{fig:asym_arch}, edge devices often face power consumption or computing resource limitations that make it impractical to deploy AE-based methods. 
A single encoded bit stream can be decoded on many different devices to mitigate the heavy resource consumption during the encoding process completed at the server side. 

However, previous INR-based methods have certain limitations. Early works achieve fast and low-complexity decoding process but failed to provide a sufficiently good rate-distortion performance \cite{dupont2021coin, dupont2022coin++, strumpler2022implicit}.
While low MACs in COOL-CHIC-like methods do not necessarily guarantee fast decoding, the primary reason is that these methods employ a pixel-by-pixel auto-regressive approach to encode and decode the latents, which is suitable for modeling the distribution of pixels but challenging to parallelize\cite{van2016pixel, van2016conditional}. This feature hinders us from using accelerator such as GPUs to improve decoding speed, which cripples the advantage of low-complexity.
Few previous studies pay attention to both quality and decoding time, which are crucial in practical applications.

In this paper, we focus on improving decoding efficiency while keep reconstruction quality. Inspired by \cite{reed2017parallel, he2021checkerboard, li2023neural},
we utilize the spatial correlation between latents with different resolution to develop a channel-wise auto-regressive causal context, which is omitted in COOL-CHIC architecture. For intra-latent encoding, we replace pixel-wise auto-regressive prediction with parallelizable checkerboard prediction.
Thorough experiments over representative datasets were performed, in which our method demonstrates superior efficiency while maintaining competitive quality and achieves higher acceleration when relax the quality requirements (Fig. \ref{fig:header_plot}). Our contributions include:
\begin{itemize}
    \item We introduce parallelization-friendly AutoRegressive upsampler (ARU) blocks, which are highly computationally efficient and employ a two-pass checkerboard strategy to enhance the utilization of context information, improving the reconstruction quality. Based on the module,  we create a novel Mixed AutoRegressive Model (MARM), whose ARU and ARM is adjustable to achieve a more flexible trade-off between quality and decoding  speed.

    \item We propose a new synthesis module capable of leveraging the inductive biases of both MLPs and CNNs, which significantly enhances the reconstruction quality compared with the original module.

    \item Extensive experiments under various parameter settings and computational environments have confirmed the effectiveness and performance characteristics of our method.

\end{itemize}

\section{Related Work}

\subsection{Neural Image Compression}
Classical neural image compression methods extend the framework of transform encoding \cite{2001Theoretical}. In this framework, both analysis transform $g_a(\boldsymbol{x};\phi_g)$ and synthesis transform $g_s(\hat{\boldsymbol{y}}; \theta_g)$ use neural network parameterized by $\phi_g$ and $\theta_g$ as transform functions, rather than linear transforms. In coding procedure, latent representation $\boldsymbol{y}$ generated by $g_a(\boldsymbol{x};\phi_g)$ is quantized to discrete
$\hat{\boldsymbol{y}}$ and losslessly compressed using entropy encoder \cite{balle2016end}.

The process of quantizing a continues $\boldsymbol{y}$ to a finite set of discrete values will bring problems of information loss and non-differentiable characteristic. The information loss leads to the rate-distortion trade-off
\begin{equation}
    \mathcal{L}_{\phi_g, \theta_g} = D+\lambda R.
\end{equation}
In training stage, the quantization is relaxed by adding standard uniform noise to make the full model differentiable
\begin{equation}
    q(\tilde{\boldsymbol{y}} | \boldsymbol{x}, \theta_g) = \prod_i\mathcal{U}(\tilde{y}_i|y_i-\frac{1}{2}, \tilde{y}_i|y_i+\frac{1}{2}).
\end{equation}
In the framework, the loss function equal to the standard negative evidence lower bound (ELBO) used in variational autoencoder (VAE) training.

There are a lot of papers follow the above framework. 
\citet{Ball2018Variational} add scale hyperprior to capture more structure information in latent representation. 
\citet{2018Joint} use an autoregressive and hierarchical context to exploit the probabilistic structure. \citet{cheng2019energy} improve bit allocation by energy compaction-based methods.
\citet{2020Channel} investigate the inter-channel relation to accelerate the encoding and decoding process. \citet{2022ELIC} use both inner-channel and inter-channel context models and improve the performance. \citet{mentzer2020high} achieve achieve high quality reconstructions by taking advantages of Generative Adversarial Networks.  Besides, some works focus on variable rate compression \cite{akbari2021learned} and scalable compression \cite{mei2021learning}, which are more practical in real scenarios.

In these methods, users have to deploy the pre-trained models on both encoding and decoding sides, which may bring problems as depicted in the previous section. But at same time, many insights proposed by these works can also apply to INR-based methods.

\subsection{Implicit Neural Representation}
Different from the end-to-end models that use real signals like images or videos as input, implicit neural representation (INR) models generally use coordinates as model input. The network itself is the compressed data representation. 
This idea thrives on 3D object representation. NeRF \cite{2020NeRF} synthesizes novel views of complex scenes by an underlying continuous volumetric scene function. The function maps the 5D vector-valued input including coordination $(x, y, z)$ and 2D viewing direction $(\theta, \phi)$ to color $(r, g, b)$ and density $\sigma$. MLP is used to approximate the mapping function. To improve model performance, positional encoding is used to enhance visual quality and hierarchical volume sampling is used to accelerate training process.

In addition to NeRF, many insights are proposed by a large body of literature. \citet{park2019deepsdf} represent shapes as a learned continuous Signed Distance Function (SDF) from partial and noisy 3D input data. 
\citet{chen2019learning} perform binary classification for point in space to identify whether the point is inside the shape.
Then the shape could be generated from the result. 
\citet{muller2022instant} proposed to use multi-resolution hash encoding to argument coordinate-based representation and achieve significant acceleration in both training and evaluation without sacrificing the quality. \citet{shao2024polarimetric} employed the implicit representation for object's geometry with a neural network to improve reconstruction quality of transparent objects.

INR is also used in image-relevant tasks. \citet{chen2021learning} extends coordinate-based representation to 2D images and develops a method that can present a picture at arbitrary resolution. \citet{2021Generative} propose to generate parameters of the implicit function instead of grid signals such as images in generative models to improve the quality.

Although INR-based methods have succeeded in many areas, popularizing of the technique in compression is non-trivial. The main difference between compression and the tasks above is the model size. In the INR-based compression method, model parameters are also part of the information that needs to be transmitted, which raises the trade-off between model size and reconstruction quality.

\subsection{INR Based image compression}

In image compression, COIN \cite{dupont2021coin} uses standard coordinate representation that directly maps 2D coordinates $(x, y)$ to color $(r, g, b)$, which allows variable resolution decoding and partial decoding. 
Along with architecture search and weight quantization to reduce the model size, COIN outperforms JPEG for low bit rate.
COIN++ \cite{dupont2022coin++} extends the idea of a generative INR method that compresses modulation rather than model weight to achieve data-agnostic compression. In some dataset, COIN++ achieve significant performance improvement.

INR-Based image codecs like COIN transform image compression task into model compression task, which extends the scope of techniques used in the domain.
COMBINER \cite{guo2023compression} uses A* relative entropy coding \cite{havasi2019minimal, flamich2022fast} to compress
the Bayesian INR of an image. RECOMBINER \cite{he2023recombiner} further improves the RD performance by using more flexible variational approximation and learnable positional encoding.

However, in universal image compression, COIN-like and COMBINER-like methods failed to compete with AE-based neural image codec \cite{Ball2018Variational} and JPEG2000 for a high bit rate. \citet{ladune2022cool} proposed COOL-CHIC that uses latent along with an autoregressive decoding process to achieve comparable RD performance to AE-based method with low complexity. \citet{leguay2023low} push the performance forward to surpass HEVC in many conditions by leveraging a learnable upsampling module and convolution-based synthesis.

One of the disadvantages of COOL-CHIC-like methods is the theoretical low complexity and slow decoding process because of highly serial decoding process. We propose to replace the ARM model in COOL-CHIC  with a parallelization-friendly one to significantly reduce the decoding time.

\section{Method}

\begin{figure*}[h!]
  \centering
  \includegraphics[width=\linewidth]{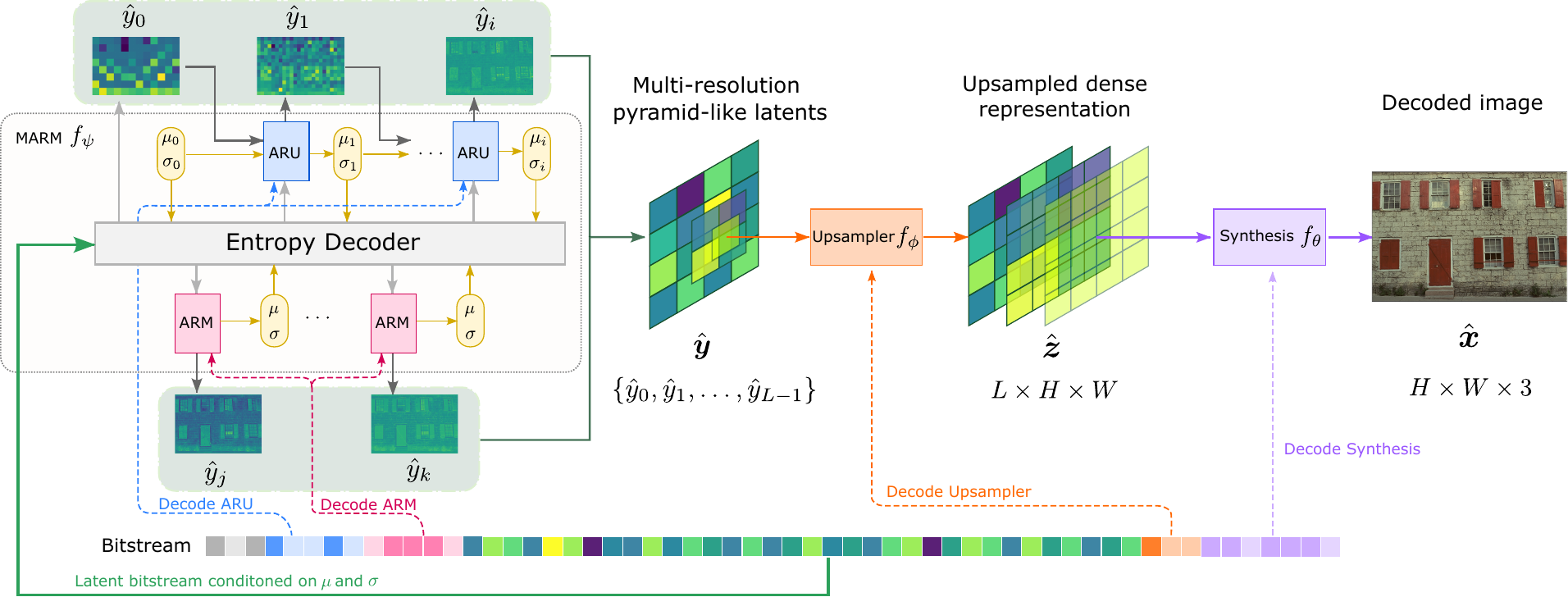}

  \caption{System architecture.  $\psi$, $\phi$, $\theta$ are network parameters for MARM module, upsample module and synthesis module respectively. These parameters are decoded first to initialize the model. Then the MARM module decodes latents $\hat{\boldsymbol{y}}$ which are integer values matrices with pyramid shapes. Upsample module will transform these latents to a dense representation $\hat{\boldsymbol{z}}$  with shape $L\times H\times W$. Synthesis module transforms the dense tensor
  to image with $(r, g, b)$ channels and $H\times W$ shape. Note since the code length of $\psi, \phi, \theta$ is vary small, we use original notation represents both quantized and unquantized version for simplicity. Top part of MARM is channel-wise auto-regressive model, which predict entropy parameters conditioned on previous latent. Bottom part correspond to inner-channel auto-regression model generation parameters pixel by pixel. Given the number of ARM blocks $M$ and total latents number $L$, we note $i=L-M-1$, $j=L-M$ and $k=L-1$. $\mu_0$ and $\sigma_0$ are initialized to tensor with 0 and $\exp(-0.5)$ respectively for all images. $\mu$ and $\sigma$ with subscription means the values output as a matrix rather than serially generated scalar for those without subscription. }
  \label{fig:arch}
\end{figure*}

\subsection{System Overview}
In image compression task, we define $\boldsymbol{x} \in\mathbb{N}^{C\times H\times W}$ as the $H\times W$ image to be compressed with $C$ channels. For common RGB pictures, $C=3$.
$\hat{\boldsymbol{x}} \in\mathbb{N}^{C\times H\times W}$ is the decoded image. As shown in Fig. \ref{fig:arch}, our model includes three modules: mixed autoregressive model $f_\psi$, upsampler $f_\phi$ and  synthesis $f_\theta$.
These networks are parameterized by $\psi$, $\phi$ and  $\theta$ respectively. 
$\hat{\boldsymbol{y}}$ is a set of pyramid-like multi-resolution latent variables with discrete values:
\begin{equation}
\hat{\boldsymbol{y}} = \left\{\hat{y_i}\in\mathbb{Z}^{H_i\times W_i}, i=0, 1, \dots, L-1\right\},\\
\end{equation}
where $H_i = \frac{H}{2^{L-i-1}}, W_i = \frac{W}{2^{L-i-1}}$. $L$ is the number of latents. Under these notations, the image is $\boldsymbol{x}$ encoded as $\{\psi, \phi, \theta, \hat{\boldsymbol{y}}\}$.

When decompressing an image, the first step is decoding the network parameters $\psi$, $\phi$ and $\theta$ and initializing the whole model. Then $\hat{\boldsymbol{y}}$ is decoded from bitstream by $f_\psi$:
\begin{equation}
  \hat{\boldsymbol{y}} = f_\psi(\boldsymbol{b}),
\end{equation}
where $\boldsymbol{b}$ represents bitstream. Like autoencoder codec, $f_\psi$ may have specific structure such as an autoregressive network \cite{leguay2023low}. Because making use of predictions from causal context of each symbol in this stage is very important to remove redundancy and reduce bit rate \cite{Ball2018Variational, 2018Joint, 2020Channel, cheng2020learned}.
After that, a dense representation $\hat{\boldsymbol{z}}\in\mathbb{R}^{L\times H\times W}$ is obtained by the learnable upsampler $f_\phi$:
\begin{equation}
  \hat{\boldsymbol{z}} = f_\phi(\hat{\boldsymbol{y}}).
\end{equation}
Finally, decoded image $\hat{\boldsymbol{x}}$ is reconstructed from $\hat{\boldsymbol{z}}$
\begin{equation}
  \hat{\boldsymbol{x}} = f_\theta(\hat{\boldsymbol{z}}).
\end{equation}

For encoding stage, different from AE-based neural image codec, implicit neural representation based neural image codec does not require encoder.
The encoding process of such methods is the process of training neural networks.
Although the coding process is different, the final target function is the same:
\begin{equation}
  \mathcal{L} = D({\boldsymbol{x}},  \hat{\boldsymbol{x}}) + \lambda R(\hat{\boldsymbol{y}}),
  \label{eq:loss_func}
\end{equation}
where $D$ is distortion function such as mean squared error or perception error and $R$ approximate rate with entropy. Since the discrete value $\hat{\boldsymbol{y}}$ is non-differentiable, which is common to deep-learning compressor, we use a set of real value $\boldsymbol{y}$ with same shape as $\hat{\boldsymbol{y}}$ and a quantization function $Q$ in training
\begin{equation}
  \hat{\boldsymbol{y}} = Q(\boldsymbol{y}).
\end{equation}
$Q$ could be either a fixed uniform scalar quantizer \cite{balle2016end} or $\epsilon$-STE quantizer \cite{leguay2023low} according to training stage.

\subsection{Mixed Autoregressive Model}
Autoregressive network is widely used in casual context prediction \cite{2018Joint, leguay2023low}, 
which demonstrates the effectiveness of the structure in reducing redundancy of compressed representation. This is more clear if we decompose the second term of Eq. \ref{eq:loss_func}
\begin{equation}
  \begin{split}
    R(\hat{\boldsymbol{y}})=D_{KL}(q||p_\psi) + H(\hat{\boldsymbol{y}}),
  \end{split}
\end{equation}
where $D_{KL}$ stands for the Kullback-Leibler divergence and $H$ for Shannon’s entropy. The first term suggest the closer we approximate to real distribution $q$, the more bit we will save. In ARM model, $p_\psi$ is decomposed as:
\begin{equation}
  p_\psi(\hat{\boldsymbol{y}}) = \prod_{l, i}p_\psi(\hat{y}_{l, i} | \mu_{l, i}, \sigma_{l, i}), 
\text{where}\,
  \mu_{l, i}, \sigma_{l, i} = f_\psi(\hat{y}_{l, < i}),
  \label{eq:arm}
\end{equation}
$l$ means the $l$-th latent and $<i$ means all pixels in a flatten latent whose index is smaller than $i$. Obviously, the decoding proceeds pixel by pixel, which is time consuming and hard to parallelize.
To alleviate such problem, our autoregressive upsampler (ARU) apply autoregressive decoding across latents.
In other word, we use low-resolution latent to predict the decoding parameter of next high-resolution latent:
\begin{equation}
  p_\psi(\hat{\boldsymbol{y}}) = \prod_{l}p_\psi(\hat{y}_l | \mu_l, \sigma_l),
\end{equation}
where $\mu_l, \sigma_l = f_\psi(\hat{y}_{l - 1}, \mu_{l-1}, \sigma_{l-1})$. This approach can significantly improve the parallelism of the autoregressive module and greatly enhance the computational performance. Similar technique is also used in some previous work \cite{reed2017parallel}.

While ARU block outperforms in efficiency, ARM can recognized more correlation between adjacent pixels because of locality inside each latent.
So we integrate ARU and ARM to a  Mixed AutoRegressive Model (MARM), as shown in Fig. \ref{fig:arch}. For low-resolution latents, which have more global information, ARU is used to accelerate decoding process. For high-resolution latents, we use ARM to capture more details such as textures. The ratio of two type blocks is controlled by a hyperparameter $M$, which means the number of ARM blocks in MARM. Note when $M=L$, the MARM becomes pure ARM context model, similar to COOL-CHIC.

\subsection{Two Pass ARU}

\begin{figure*}[ht]
  \subfloat[ARU architecture details]
  {\includegraphics[width = 0.64\textwidth]{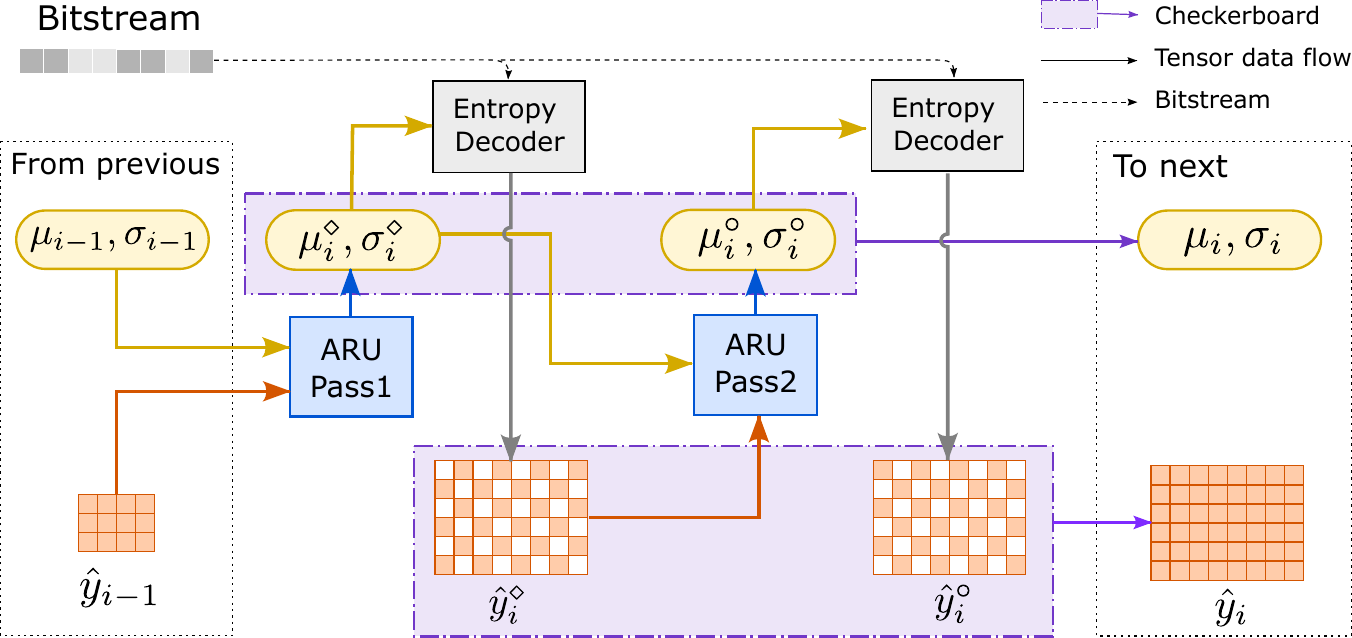}\label{fig:aru_details_sub}} \hfill
  \subfloat[Visualization of latent in checkerboard decoding]{\includegraphics[width = 0.34\textwidth]{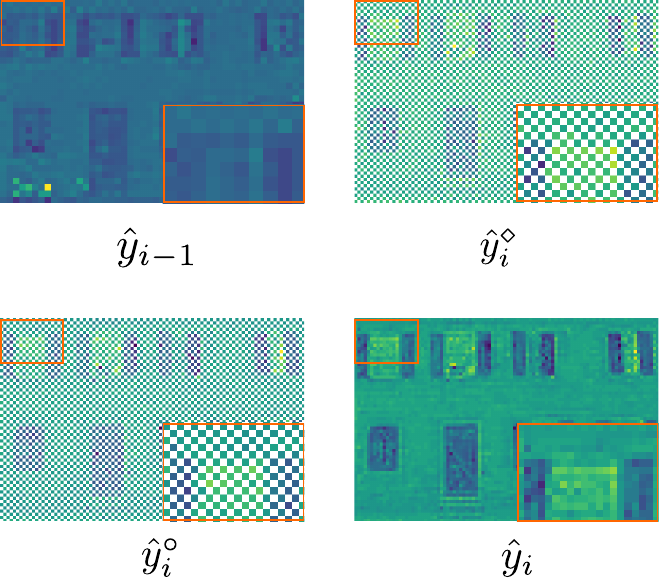}\label{fig:latent_vis}}

  \caption{Network architecture of ARU (Fig. \ref{fig:aru_details_sub}). ARU uses previous parameter matrix $\mu_{i-1}$, $\sigma_{i-1}$ and latents as input. ARU pass1 output decoding parameters $\mu_i^{\diamond},  \sigma_i^{\diamond}$ for anchors in latents. ARU Pass2 generate parameters for non-anchors conditioned on $\mu_i^{\diamond},  \sigma_i^{\diamond}$ and anchors.  Checkerboard means to merge anchor of the first tensor and non-anchor of the second tensor. Fig. \ref{fig:latent_vis} provides an example of latent in decoding process.}
  \label{fig:aru_details}
\end{figure*} 

Although using low-resolution latent to predict higher-resolution ones is target to improve computational performance at the cost of reconstruction quality, the degradation can be reduced. Different from pixel-by-pixel correlation or cross resolution correlation, we can utilize the locality in only two pass in a checkerboard fashion. As shown in Fig. \ref{fig:aru_details}, we mark anchor in tensor $\hat{y}_i$,  $\mu_i$,  $\sigma_i$ (orange ones of $\hat{y}_i^{\diamond}$ in Fig.\ref{fig:aru_details} ) as $\hat{y}_i^{\diamond}$,  $\mu_i^{\diamond}$,  $\sigma_i^{\diamond}$, 
non-anchor (white ones of $\hat{y}_i^{\diamond}$ in Fig.\ref{fig:aru_details} ) as $\hat{y}_i^{\circ}$, $\mu_i^{\circ}$,  $\sigma_i^{\circ}$ respectively. 
Following the notation, joint distribution of $\hat{y}_i$ can be written as
\begin{equation}
\begin{split}
 p_\psi(\hat{y}_i | \mu_i, \sigma_i ) 
    &= p_\psi(\hat{y}_i^{\diamond}| \mu_i, \sigma_i) \cdot p_\psi(\hat{y}_i^{\circ} | \hat{y}_i^{\diamond}, \mu_i, \sigma_i)\\
    &= p_\psi(\hat{y}_i^{\diamond} | \mu_i^{\diamond},  \sigma_i^{\diamond}) \cdot p_\psi(\hat{y}_i^{\circ} | \mu_i^{\circ},  \sigma_i^{\circ}).   
\end{split}
\end{equation}

The anchor pixels only depend on information from previous low-resolution latent, and the correlation is fitted by $f_\psi^\diamond$:
\begin{equation}
  \mu_i^{\diamond},  \sigma_i^{\diamond} = f_\psi^\diamond(\hat{y}_{i-1}, \mu_{i-1}, \sigma_{i-1}).
  \label{eq:aru_pass1}
\end{equation}
For decoding of non-anchor pixels, all previous information is available, including the decoded value of anchor $\hat{y}_i^{\diamond}$. As another form of making use of causal context information, ARU Pass2 can compute $\mu_i^{\circ}$ and $\sigma_i^{\circ}$ accordingly
\begin{equation}
  \mu_i^{\circ},  \sigma_i^{\circ} = f_\psi^\circ(\hat{y}_i^{\diamond}, \mu_i^{\diamond},  \sigma_i^{\diamond}).
\end{equation}

The idea of parallel predicting probability mass function of compressed representations have been investigated in some AE-based image or video codec \cite{he2021checkerboard, li2023neural}. Same as these methods, INR codec also benefits from this design.

\begin{figure*}
  \subfloat[ARU Pass 1 network]
  {\includegraphics[height = 0.33\textwidth]{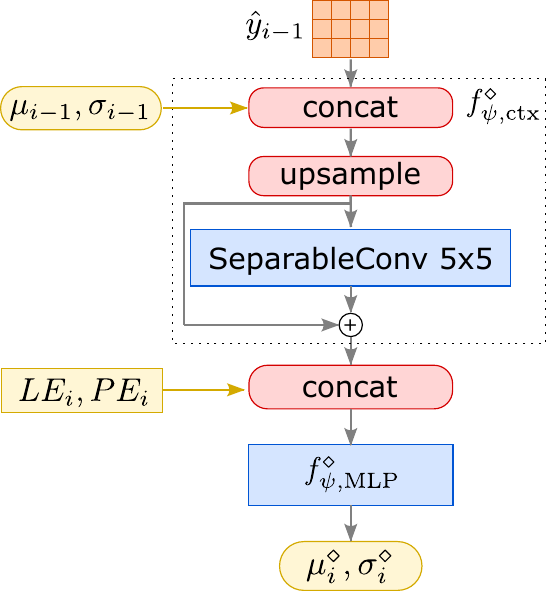}
  \label{fig:aru_pass1}} \hfill
  \subfloat[ARU Pass 2 network]
  {\includegraphics[height = 0.33\textwidth]{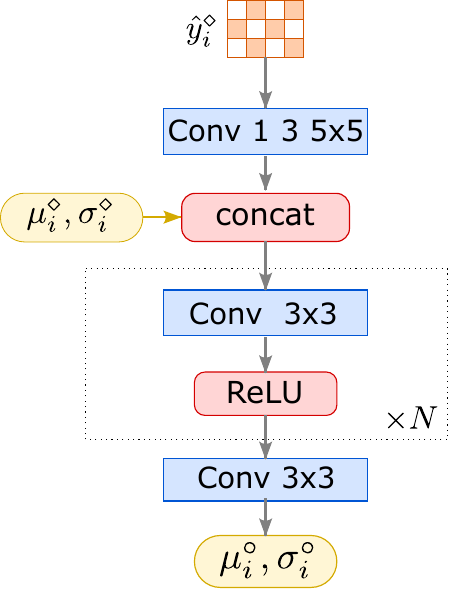}
  \label{fig:aru_pass2}} \hfill
  \subfloat[Mixed synthesis module]
  {\includegraphics[height = 0.33\textwidth]{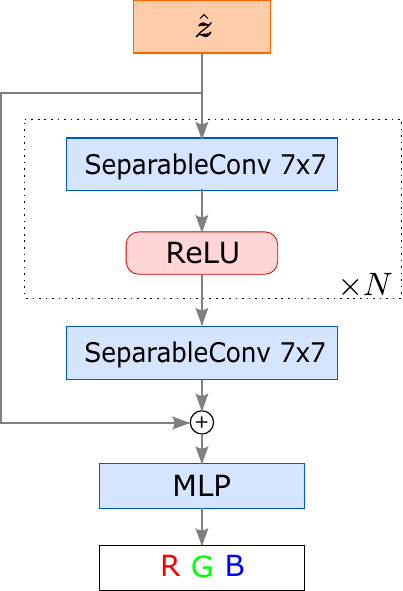}\label{fig:sythesis}} \hfill
  
\caption{Detailed structure of ARU and Synthesis network. The input and output channel of all Conv module are defined according to their input tensor. For example, if we use latents $L=7$, $\hat{\boldsymbol{z}}$ in Fig. \ref{fig:sythesis} is a $1\times L\times H \times W$ tensor. Then input and output channel of SeparableConv is 7.}
\label{fig:network_details}
\end{figure*}

The detailed structure of two ARU sub-blocks is show in Fig. \ref{fig:aru_pass1} and \ref{fig:aru_pass2}. 
For simplicity, we only introduce the overall data flow of the module. In addition to previous content, we add level encoding and positional encoding to promote the networks capability. As shown in Fig. \ref{fig:aru_pass1}, Eq. \ref{eq:aru_pass1} can be decomposed to three steps:
 \begin{equation}
  v_{\rm{ctx}} =  f_{\psi, \rm{ctx}}^\diamond(\hat{y}_{i-1}, \mu_{i-1}, \sigma_{i-1}),
\end{equation}
\begin{equation}
  v_{i, ab}^\diamond =  {\rm Concat}([v_{\rm{ctx}, ab}, PE_{i,ab}, LE_{i, ab}]),
\end{equation}
\begin{equation}
  \mu^{\diamond}_{i, ab},  \sigma^{\diamond}_{i, ab} =  f_{\psi, \rm{MLP}}^\diamond(v_{i, ab}^\diamond).
 \end{equation}
 
To reduce the dimension of the output, we only use a simple mesh grid positional encoding.
Let $\{(a, b)\in\mathbb{Z}^2, 0\leq a<H_i, 0\leq b < W_i-1\}$ represents spatial location in $i$-th latent, the position encoding is defined as
\begin{equation}
  PE_{i,ab} =  [\frac{a}{H_i} - 0.5, \frac{b}{W_i} - 0.5].
 \end{equation}
And the level encoding assigned to each latent is defined as
 \begin{equation}
  LE_{i, ab} =  \frac{2i}{L}.
 \end{equation}
Level encoding and positional encoding provide more information to ARU networks.
\subsection{Mixed Synthesis Module}
Previous work have investigated the performance of full MLP \cite{ladune2022cool} and full convolutional network \cite{leguay2023low} as
synthesis. However the prior enforced by both structure may not apply for all input data. 
To enhance the generality of method, we combine MLP and convolution layer by a residual connection, which cooperates the information of adjacent latents pixels and increases the network depth, as shown in Fig. \ref{fig:sythesis}.

We use Separable Convolution module \cite{chollet2017xception, howard2017mobilenets} in Synthesis to reduce the parameters number while maintaining model capacity as possible. Kernel size of convolution layer is 7 to increase reception field. The number of convolution layer $N$ is 3. The width of MLP is 12 with 2 hidden layers. 
 This design further improves the reconstruction quality.

\subsection{Complexity Analysis}
In COOL-CHIC-like methods, the process of decoding latent takes part majority of decoding time in many cases. Given an image with $n=H\times W$ pixels, the total number latent pixels need to be decoded is 
\begin{equation}
\begin{split}
   N(n)
    &= \sum_{i=0}^{L-1} c H_i W_i = HW\sum_{i=0}^{L-1} \frac{c}{2^{2(L-i-1)}} \\
    &=O(n).
\end{split}
\end{equation}
Because of serial decoding, time complexity of decoding latent is the same. Considering the computation time of neural network is proportional to total pixels $n$ as well, the time complexity of overall decoding process is $O(n)$. 

Previous work introduce wave-front decoding to accelerate the process \cite{ladune2022cool, leguay2023low}, which can improve the constant factor in complexity. Another acceleration method is decoding all latents in parallel since these latent are independently encoded. However, the decoding time of the largest latents still dominate the time consumption, which makes the improvement marginal.

Modern hardware like GPU is good at handling massively parallel computations. Our design of ARU can exploit the benefits.
If we suppose parallel operations such as convolution operation over a feature map including a smaller number channels ($<10$) can be finished at constant time $c_i$, which is practical for common size pictures on modern GPU,
the decoding time complexity of MARM is 
\begin{equation}
\begin{split}
   T(n)
    &= T_{\text{ARUs}}(n) + T_{\text{ARMs}}(n) + T_{\text{NNs}}(n) \\
    &=\sum_{i=0}^{L-M-1}c_i + \sum_{i=L-M}^{L-1} c H_i W_i + c_\text{NNs} \\
    &=O(Mn)
\end{split}
\end{equation}

When $M=0$, the complexity of our method becomes $O(1)$, which surpass the previous work.
When $M > 0$, the complexity is similar to previous work. But in realistic setting, $n$ is finite, which means the constant factor is important as well. Actually, experiments support when $M\leq2$, the acceleration is still significant. For device without GPU, our method can outperform previous work in speed because most of modern CPUs support SIMD (x86) or NEON (ARM), which can also accelerate parallel computations.

\section{Experiment}

\subsection{Datasets, Metrics and Experiments Setup}
\label{sec:exp:subsec1}
The experiment use images from Kodak dataset\footnote{http://r0k.us/graphics/kodak} and CLIC professional valid set\footnote{https://clic.compression.cc/2021/tasks/index.html}.
Kodak dataset is a widely used dataset in image compression community, which includes 24 images of size $768\times 512$. The main experiment results and more ablations are performed in this dataset.
CLIC dataset contains a collections of natural images with different resolution. 
Since training set in unnecessary for INR codecs, we only perform experiments on valid set. 

\begin{figure*}
\subfloat[TD curve of Kodak]
  {\includegraphics[width=0.49\linewidth, trim={0 0 1cm 1cm },clip]{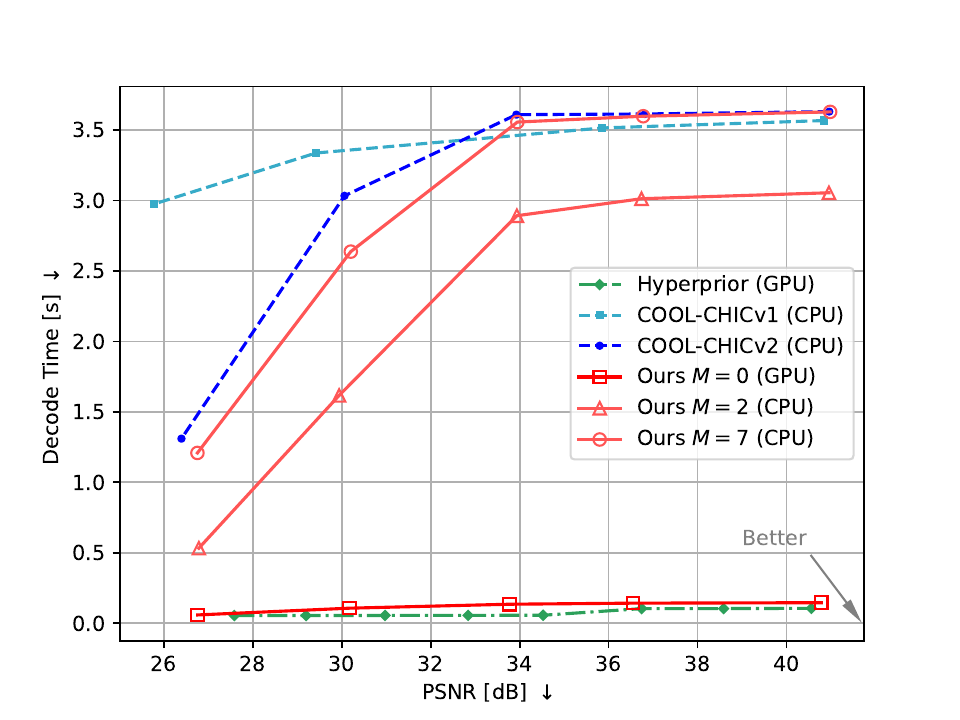}
\label{fig:kodak_psnr_time}
} \hfill
\subfloat[TD curve of CLIC]
  {\includegraphics[width=0.49\linewidth, trim={0 0 1cm 1cm },clip]{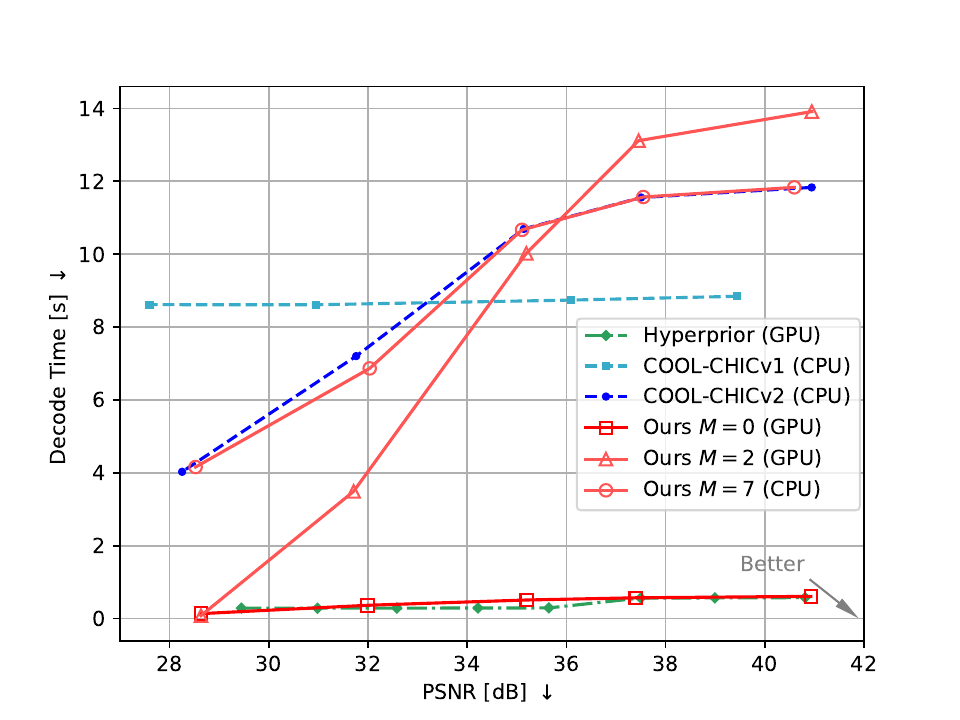}
        \label{fig:clic_psnr_time}
}
\caption{Decoding time results. Fig. \ref{fig:kodak_psnr_time} and Fig. \ref{fig:clic_psnr_time} show the time-distortion performance averaged over Kodak dataset and  CLIC professional valid dataset respectively. We only draw the best result among different settings (marked in brackets) for each method for simplicity. Full results are shown in Table \ref{table:main_results}. }
\label{fig:main_time}
\end{figure*}

We use the peak signal-to-noise ratio (PSNR) in RGB 4:4:4 as quality metric, which is the most widely used metric in image compression task.  Bit-per-pixel (BPP) is used as coding efficiency metric.
To achieve comprehensive evaluation across different qualities, we report Bjøntegaard-Delta-Rate (BD-Rate) \cite{Bjntegaard2001CalculationOA} of each method. We also extends rate-distortion curve to time-distortion curve (TD curve), which consists of $(T_\text{decoding time}, D)$ pairs (Fig. \ref{fig:main_time}). TD curve effectively illustrates the decoding speed across various reconstruction quality levels. Similar to BD-Rate, we propose using BD-Time, which measure the average difference between two TD curves, to evaluate the performance in terms of decoding speed. For calculation, we can simply replace the bit rate term with decoding time in BD-Rate formula to obtain the value of BD-Time.

To ensure fairness in  comparison, all learning-based model is implemented using PyTorch without special optimization. For AE-based models, we use the pre-trained models in CompressAI \cite{Jean2020CompressAI}.
Our model is implemented based on previous works \cite{ladune2022cool, leguay2023low}, which use constriction package \cite{bamler2022constriction} as entropy encoder. For INR-based codec, we only use COOL-CHIC-like methods \cite{ladune2022cool, leguay2023low} as baselines since other methods do not have competitive reconstruction quality especially in high bitrate settings\cite{dupont2021coin, guo2023compression, he2023recombiner}. We follow the original settings for all baseline methods. We use $\lambda = \{0.02, 0.004, 0.001, 0.0004, 0.0001\}$ to evaluate the quality on different bit-rate. At training stage, we use learning rate decreasing strategy same as COOL-CHICv2.
MSE is used as distortion function for all training based methods. 

Experiments are conducted on both CPU and GPU to facilitate an assessment aligned with real-world scenarios. We use the same server (2.20 GHz, 10C/20T) with single GPU (14.2 TFLOPS peak FP32 performance) to benchmark all methods. When decoding on CPU, the number of CPU cores is limited to one unless otherwise specified. The reason is twofold. Firstly, using single-core decoding simulates some resource-constrained decoding scenarios like AR
devices. Secondly, some baseline methods can not effectively utilize multi-core parallelism. For methods that are parallelism-friendly, results on GPU are more indicative. We provide more ablations about the setting in Section \ref{sec:experiments:ablation}. 

\begin{figure*}
\subfloat[RD curve of Kodak]
  {\includegraphics[width=0.49\linewidth, trim={0 0 1cm 1cm },clip]{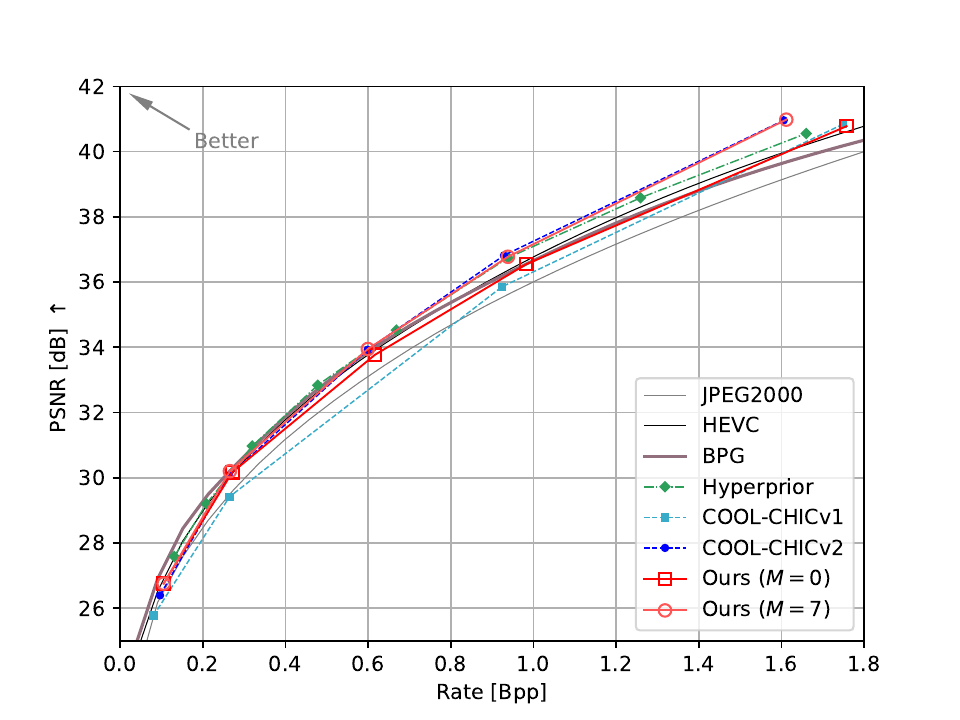}
\label{fig:kodak_bpp_psnr}
} \hfill
\subfloat[RD curve of CLIC]
  {\includegraphics[width=0.49\linewidth, trim={0 0 1cm 1cm },clip]{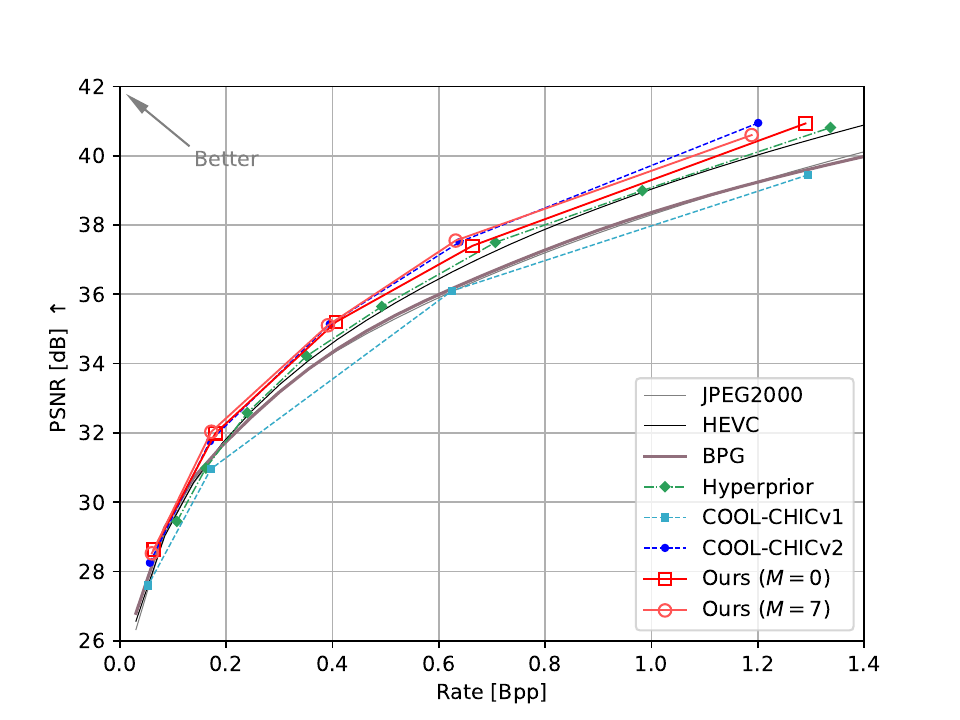}
        \label{fig:clic_bpp_psnr}
}
\caption{Reconstruction quality results. Fig. \ref{fig:kodak_bpp_psnr} and Fig. \ref{fig:clic_bpp_psnr} show the rate-distortion performance averaged over Kodak dataset and  CLIC professional valid dataset respectively.
}
\label{fig:main_result}
\end{figure*}

\subsection{Image Compression}

\begin{figure*}
  \subfloat[Ground truth]
  {\includegraphics[width = 0.166\textwidth]{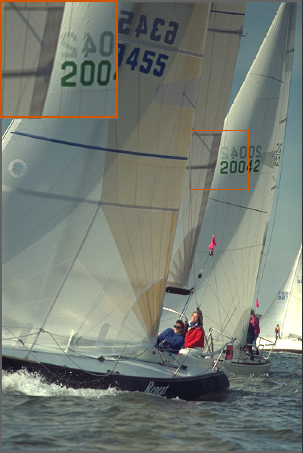}} \hfill
  \subfloat[CCv1, 0.0005\\(0.573, 3.317s, 37.035)]{\includegraphics[width = 0.166\textwidth]{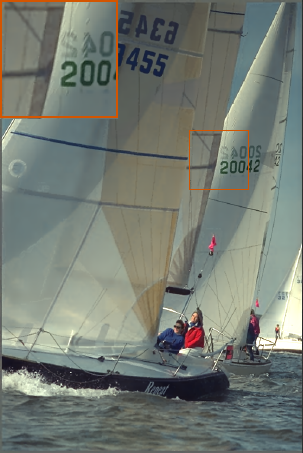}} \hfill
  \subfloat[CCv2, 0.0004\\(0.559, 4.229s, 37.560)]{\includegraphics[width = 0.166\textwidth]{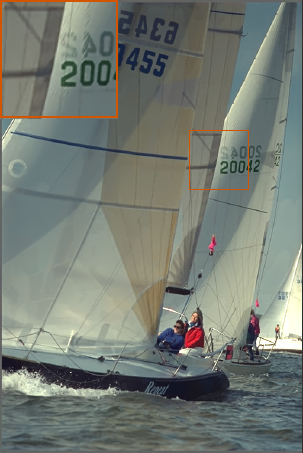}} \hfill
  \subfloat[Ours $M=0$, 0.0004\\(0.583, \textbf{0.136s}, 37.504)]{\includegraphics[width = 0.166\textwidth]{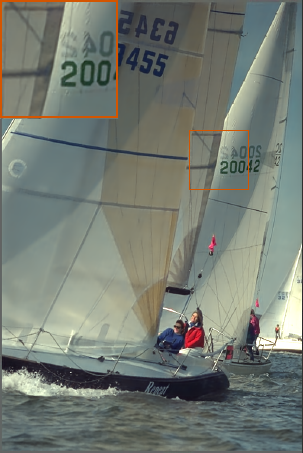}} \hfill
  \subfloat[Ours $M=2$, 0.0004\\(0.560, 3.284s, 37.785)]{\includegraphics[width = 0.166\textwidth]{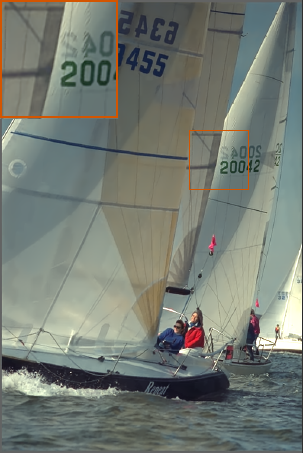}} \hfill
  \subfloat[Ours $M=7$, 0.0004\\(\textbf{0.553}, 3.983s, \textbf{37.889})]{\includegraphics[width = 0.166\textwidth]{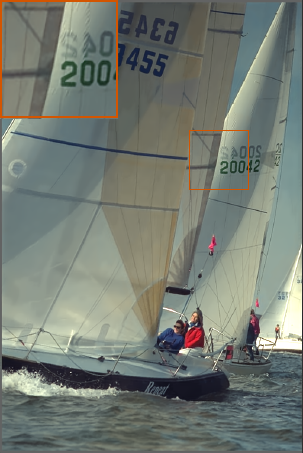}}  \\
  
  \hspace{0.166\textwidth}\hfill
  \subfloat[CCv1, 0.004\\(\textbf{0.157}, 3.352s, 31.126)]{\includegraphics[width = 0.166\textwidth]{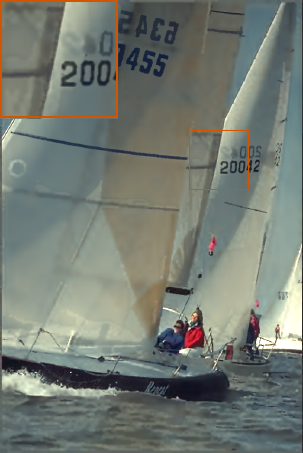}}
	\hfill
 \subfloat[CCv2, 0.004\\(0.168, 3.986s, 31.916)]{\includegraphics[width = 0.166\textwidth]{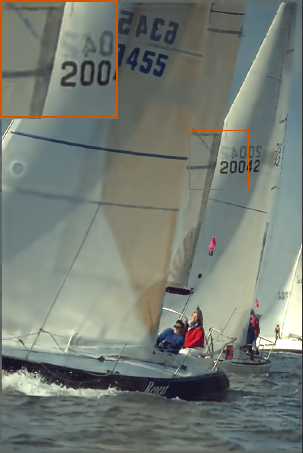}}	\hfill
 \subfloat[Ours $M=0$, 0.004\\(0.176, \textbf{0.112s}, \textbf{32.110})]{\includegraphics[width = 0.166\textwidth]{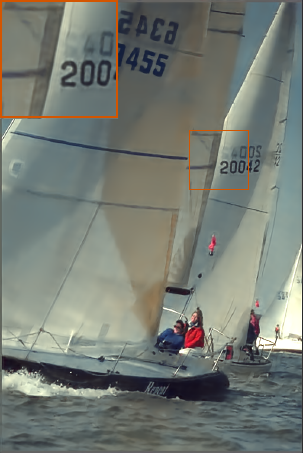}}	\hfill
 \subfloat[Ours $M=2$, 0.004\\(0.173, 1.256s, 31.171,)]{\includegraphics[width = 0.166\textwidth]{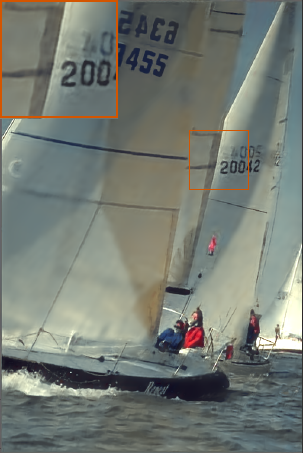}}	\hfill
 \subfloat[Ours $M=7$, 0.004\\(0.172, 3.990s, 31.188)]{\includegraphics[width = 0.166\textwidth]{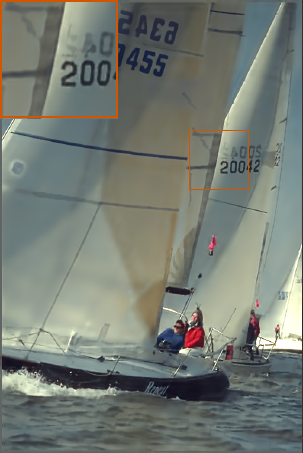}}
\caption{Visualization of decoded images. The ground truth image is kodim10.png, form Kodak dataset. COOL-CHIC is abbreviated as CC. We summarize the numerical results in subcaption of each image as \textit{Name}, $\lambda$, (\textit{BPP}, \textit{Decode Time}, \textit{PSNR}).}
\label{fig:decoded_images}
\end{figure*}

The compression performance of proposed algorithm model is compared with JPEG2000, HEVC, BPG, Hyperprior with scale prior (Hyperprior)\cite{Ball2018Variational}, COOL-CHICv1\cite{ladune2022cool} and COOL-CHICv2\cite{leguay2023low}.
Fig. \ref{fig:main_result} and  illustrates the decompression quality results of our method. Fig. \ref{fig:main_time} shows the decoding performance. More numeric results can be found in Table \ref{table:main_results}. All COOL-CHIC-like methods aggregate the results over same $\lambda$ respectively.

We observe our method achieves significant acceleration in decoding time on GPU comparing with previous COOL-CHIC-like method (about $10\times\sim 20\times$ speedup for different $\lambda$) while maintain comparable reconstruction quality ( about $1\%  \sim 5\%$ worse BD-Rate compare with COOL-CHICv2) when $M=0$ \emph{i.e.} using pure ARU blocks in MARM module. If using larger $M$, the decoding time increases but reconstruction quality improves. Note our method outperform COOL-CHICv2 in decoding speed in all settings. When $M=7$, we achieves the state-of-the-art quality comparing with other INR-based codec. 

Hyperprior has best decoding performance on GPU in our experiments, but the performance highly rely on high performance GPU since $\sim 50\times$ more MACs is required to decode same image comparing with COOL-CHIC-like methods. For many low power devices like cell phone and AR glasses, this method is too resource-intensive to be deployed. Other AE-based codec also grapple with the same challenge. Fig. \ref{fig:header_plot} provide an intuitive comparison of AE-based methods and INR-based methods. Although Hyperprior has achieved the fastest decoding speed, our method ($M=0$) is not far behind ($<2\%$ worse BD-Time) and offers superior RD performance on high resolution dataset while maintaining significantly lower complexity.

Fig. \ref{fig:decoded_images} demonstrates some visualization of decoded images. We can see all methods produce better results when $\lambda$ is small and lower bitrate when using larger $\lambda$. We show a split with number and edges to evaluate visual quality. The results in first line are close to the ground truth image. In low bitrate, all image exhibits similar distortion like color shift, which is a little different from block artifacts in traditional codec. The reason of generating this type of distortion is still an open problem.

In summary, by using MARM module and mixed synthesis, our method achieves better trade-off among RD performance, TD performance and complexity than existing methods.

\begin{table*}[ht]

   \centering
   \caption{Numerical results of our method. Both BD-Rate and BD-Time (abbreviated as BD-T) are calculated relative to COOL-CHICv1\cite{ladune2022cool} decoding on CPU. FLOPs is normalized by pixels.}
   \label{tab4}
   \begin{tabular}{c c c c c c c c c}\toprule
	\multirow{2}*{Method} & \multirow{2}*{Params} & \multirow{2}*{FLOPs } & \multicolumn{3}{c}{Kodak} & \multicolumn{3}{c}{CLIC} 	\\ \cmidrule(r){4-6} \cmidrule(r){7-9}
     ~ & ~ & ~  & BD-Rate $\downarrow$ & BD-T/CPU $\downarrow$ & BD-T/GPU $\downarrow$ & BD-Rate $\downarrow$& BD-T/CPU $\downarrow$ & BD-T/GPU $\downarrow$\\ \hline
    Hyperprior\cite{Ball2018Variational} & $5.90\times 10^6$ & $1.041\times 10^5$   & -11.9545 & \underline{-56.21} & \textbf{-97.96}   & -11.7311 & 2.27  & \textbf{-95.98} \\
    COOL-CHICv2\cite{leguay2023low} & $5.18\times 10^5$& $2.015\times 10^3$   & \underline{-12.3368} & -9.14   & 51.64    & \underline{-22.2834} & -2.94   & 9.44\\
    \hline
    Ours ($M=0$) & $5.72\times 10^5$ & $2.481\times 10^3$ & -7.7640 & \textbf{-66.53}  & \underline{-96.51}   & -21.0300 & \textbf{-35.43} &  \underline{-95.39}\\
    Ours ($M=2$) & $5.28\times 10^5$ & $2.864\times 10^3$ & -8.3302 & -36.87  & -17.59  & -21.1839 & \underline{-20.46}  & -48.27\\
    Ours ($M=7$) & $5.27\times 10^5$ & $2.889\times 10^3$ & \textbf{-12.9300} & -13.64 & 37.95  & \textbf{-24.6662} & -4.49   & 3.31\\
    \bottomrule
    \label{table:main_results}
\end{tabular}
\vspace{-0.3 cm}
\end{table*}

\subsection{Implicit Neural Representation}

\begin{table}[ht]
\centering
   \caption{Image representation result. All results are measured in PSNR }
   \label{tab4}
   \begin{tabular}{c c c  c c}\toprule
	\multirow{2}*{Method}     & \multicolumn{2}{c}{Kodak} & \multicolumn{2}{c}{CLIC}  \\ 
 \cmidrule(r){2-3} \cmidrule(r){4-5}
 ~ & w/o quant & w/ quant & w/o quant & w/ quant  \\ \hline
    COOL-CHICv1  & 50.152 & 49.205 & 48.026 & 47.189 \\
    COOL-CHICv2  & 50.875 & 47.808 & 46.854 & 45.069 \\
    Ours         & \textbf{52.413} & \textbf{52.238} & \textbf{49.209} & \textbf{49.272} \\
    \bottomrule
    \label{table:inr_result}
\end{tabular}
\end{table}
Many works focus on representing image by an INR network. INR-based compression is essentially a special neural representation problem with a bitrate constraint. Therefore, the expressive capacity of the network is important when no rate constraint is enforced. 

Similar to other INR models, the architecture of COOL-CHIC-like methods have a strong capability to approximate 2D images. To investigate the extreme potential to represent an image, we train all COOL-CHIC-like methods without rate constraint, as show in Table \ref{table:inr_result}. Different from other INR methods, if we consider the parameters contained in the latents, the total parameters of  COOL-CHIC-like methods are usually much larger than those of a typical INR method. Therefore, we only compare our method with previous COOL-CHIC-like methods. Since quantization of the latents has a significant impact on the results, we report both with and without quantization.

In implicit neural representation task, we excluded the effects of the entropy parameter estimation network \emph{i.e.} ARM or MARM module to focus solely on the performance of the synthesis network. The experiments demonstrate that the propoesed mixed synthesis network exhibits superior reconstruction quality and robustness to quantization.

\subsection{Ablation Studies}
\label{sec:experiments:ablation}
Besides the neural representation task, we conduct more ablation studies on Kodak dataset to evaluate the performance of different components of the proposed framework.

\begin{table}[t]
\centering
   \caption{Ablation results of the rate-distortion curve average over kodak dataset with different settings when $M=0$. }
   \label{tab4}
   \begin{tabular}{c c c c}\toprule
	Module        & \multicolumn{3}{c}{Settings}  \\ \hline
    Checkerboard  &  \checkmark  & \checkmark  &               \\
    Mixed Synthesis &  \checkmark  &             & \checkmark     \\ \hline
    BD-Rate $\downarrow$ & 0.0000 & 7.5453 & 7.1030 \\
    \bottomrule
    \label{table:ablation}
\end{tabular}
\end{table}

\emph{1) Checkerboard mechanism and mixed synthesis: } Table.\ref{table:ablation} illustrates the performance of checkerboard mechanism in ARU module and our proposed new synthesis module. No checkerboard means we omit the second pass when decoding latent \emph{i.e.} use $\mu_i^{\diamond},  \sigma_i^{\diamond}$ directly to decode both anchors and non-anchors in each latent.
No mixed synthesis means we use the original one in COOL-CHICv2.
It is obvious that these two structures further improve the RD performance. Combining the results from the Table \ref{table:main_results}, it should also be noted that when $M=0$, our entropy parameter estimation network still exhibits a certain gap compared to pure ARM network ($M=7$). The new mixed synthesis network has compensated for this issue to some extent.

\begin{figure}[t]
\centering
		\centering
        \includegraphics[width=\linewidth, trim={0 0 0 0 },clip]{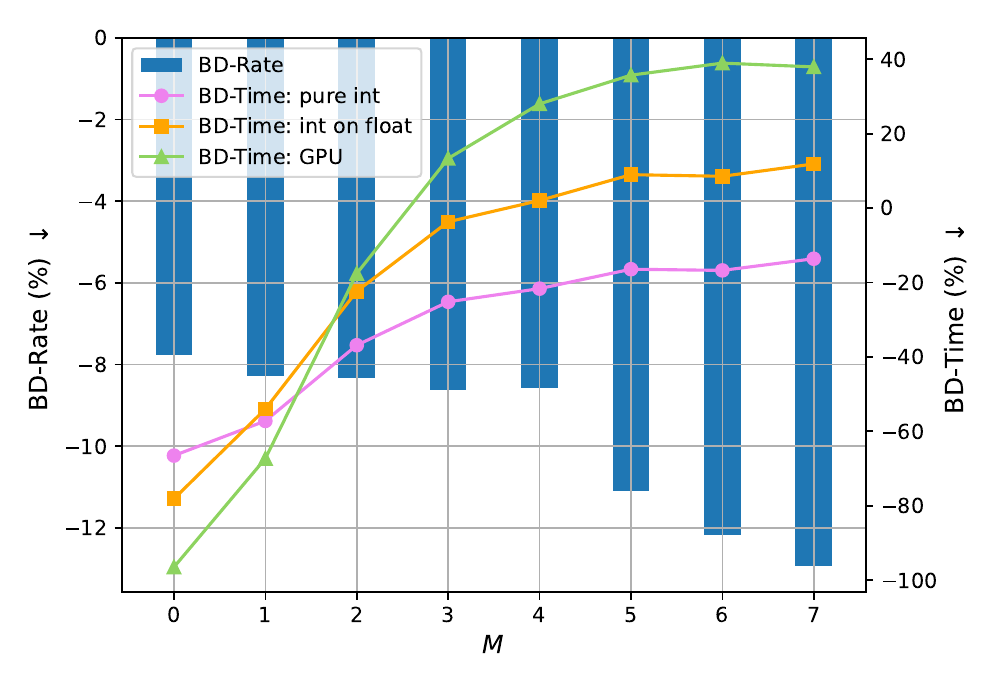}
        \caption{Quality and efficiency of different $M$. Both BD-Rate and BD-Time are computed relative to COOL-CHICv1 on Kodak dataset.}
        \label{fig:q_and_e}
\end{figure}

\emph{2) Impact of $M$ in different settings: } As mentioned before, $M$ controls the ratio of ARM blocks, which is highly correlated to the model performance. Fig. \ref{fig:q_and_e} illustrates the overall decoding quality and efficiency in different settings. Pure int mode means that all parameters in entropy network is truncated as integer after fixed scaling. During the inference \emph{i.e.} decompression, results of each layer is truncated as well. Int on float mode means to round the parameter and results while keeping the values in float.
 Both of the modes are calculated on CPU in previous work \cite{leguay2023low}.
GPU mode is int on float mode but calculated on GPU. All of the modes output image with same quality, users may choose best one according to parameters setting and platform.

The trend is easy to understand. More ARU blocks \emph{i.e.} smaller $M$  leads to more significant acceleration.
When $M$ increases, decoding time increases while performance improves. When $M=7$, our model achieve state-of-the-art RD performance. In addition to the overall trend, our method exhibits significant performance variations under different settings. When $M=0$ GPU mode outperforms others but slower than others when $M\ge2$. This suggests that the ARU block can significantly enhance the parallelism of the model, leading to substantial acceleration on GPUs. Improvements in other modes also benefit from this as well. When $M$ is large, more parameters in latents are decoded in pixel-by-pixel fashion, deteriorates the decoding performance especially on GPU.

\begin{figure}[t]
\centering
		\centering
        \includegraphics[width=\linewidth, trim={0 0 0 0 },clip]{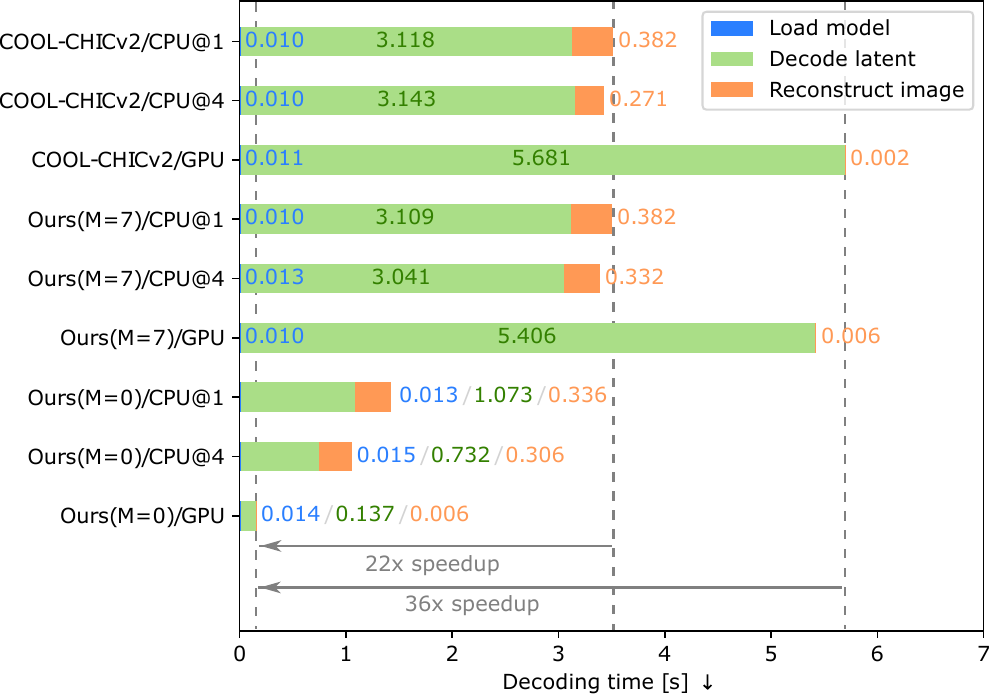}
        \caption{Time composition. We decode the kodim01.png at $\lambda=0.0001$ for each method. We mark the detail time with correspond color for clarity. As a reference, the results of bitrate (bpp) and PSNR are (COOL-CHICv2, 2.32, 40.71), (Ours $M=7$, 2.28, 40.63), (Ours $M=0$,  2.55, 40.65)}
        \label{fig:time_comp}
\end{figure}

\emph{3) Time composition: } In order to present a more detailed illustrations of performance across different settings,  we depict the detailed decoding time composition in Fig. \ref{fig:time_comp}. We show the results on CPU with 1 core (CPU@1), 4 cores (CPU@4) and on GPU.

Obviously, for all parameters setting,
the decoding time of latents dominates the decoding efficiency, which is the main focus of our work. For COOL-CHICv2 and our method when $M=7$, because of serial decoding, more CPU cores or GPU cannot accelerate the decompression process. Since entropy coder works on CPU, frequently communication between CPU and GPU further deteriorates the decoding performance on GPU, although synthesis consumes far less time than on the CPU.

This result also demonstrates the effectiveness of our improvements. When $M=0$, our method significantly outperforms others on all three settings.

\section{Conclusion}
This paper introduces a new module, MARM, designed to augment the current implicit neural representation image codec. Impressively, this is the first method to attain comparable performance in both reconstruction quality and decoding time. The MARM module enhances computational efficiency by utilizing the channel-wise autoregressive architecture in low-resolution latent and pixel-wise autoregressive mechanisms, ensuring the preservation of decompression quality. Our experimental results demonstrate that the alterations made result in a significant reduction in decoding time, without triggering substantial quality degradation.

\footnotesize{
\bibliographystyle{IEEEtranN}

}

\end{document}